\documentstyle[12pt,twoside,fleqn,espcrc1]{article}

\newcommand{\AmS}{{\protect\the\textfont2
  A\kern-.1667em\lower.5ex\hbox{M}\kern-.125emS}}

\input{epsf}
\title{E864:  Experimental Results on Exotic Nuclei and Rare Probes}

\author{J.L. Nagle\address{Department of Physics, Columbia University\\
New York, NY 10027, USA\\E-mail: nagle@nevis1.nevis.columbia.edu} for the E864 Collaboration}

\begin{document}
\maketitle

\begin{abstract}
Experiment 864 at the BNL-AGS was designed to search for exotic states
of strange quark matter and other rare composite particles.  The experiment
was commissioned in 1994 and completed its final run in 1998.  Here, we present
an overview of the experimental results for the production of
light nuclei up to $A=7$, antiprotons and antideuterons, hypernuclei and 
production limits on new states of quark matter.
\end{abstract}

\vspace{-0.05in}
\section{INTRODUCTION}

Shortly after the Big Bang the universe had expanded and cooled such that temperatures
were low enough for light nuclei (deuterons, $^{3}He$, tritons, etc.) to
coalesce.  The present abundance of these states yields significant information confirming
our picture of the evolution of the universe from its earliest stages.  In a 
similar manner, we can gain important information about relativistic heavy ion
collisions through the study of ``little bang'' nucleosynthesis.  

For over a decade, the source dimensions and flow profile of the dense nuclear
matter formed in heavy ion collisions has been studied through the measurement
of particle correlations~\cite{hbt_review}.  
The multiplicities of light nuclei
have similar information since they result from correlations between nucleons when
the hadronic system freezes out.   In addition, in heavy ion collisions, there
is a substantial abundance of antibaryons and strange baryons which may 
produce more exotic antinuclei and hypernuclei.  We present here preliminary 
results on the yields of these states and production upper limits on meta-stable
strange quark matter states.

\vspace{-0.05in}
\section{EXPERIMENT}

Experiment 864 is a two dipole, open-geometry, magnetic spectrometer.  We have 
beam counters to indicate ``good'' beam on target and give a start time for the interaction.
Particles within our acceptance pass through two large aperture dipole magnets and
are tracked through a series of time-of-flight hodoscopes and straw tube stations.  
From the track trajectory we determine the particle momentum, and together with the 
time information from the hodoscopes, we calculate the particle mass. 
For the high sensitivity searches even small cross section scattering processes will
occur and produce incorrect mass information.  A hadronic calorimeter at
the end of the spectrometer provides an independent measure of the particle mass through
the particle's energy and time-of-flight.  This detector is also read-out and used for our
``late-energy'' trigger which effectively selects events in which a high mass candidate is
present.  Since antibaryons leave additional annihilation energy in the calorimeter,
the trigger is also used to enhance the antiproton and antideuteron sample.  The experiment
in total recorded of order one billion events, which effectively sampled on the 
order of 50 billion 10\% most central $Au-Pb/Pt$ collisions.

\vspace{-0.05in}
\section{LIGHT NUCLEI}

We have measured the transverse momentum spectra for nucleons (protons and neutrons)
and light nuclei up to baryon number A=3 at $y=2.3$.   
The spectra show an increasing Boltzmann 
temperature parameter with increasing mass (proton $T=212 \pm 16~MeV$, neutron 
$T=223 \pm 23~MeV$, 
deuteron $T=347 \pm 17~MeV$, $^{3}He$~$T=405 \pm 24~MeV$).  
This feature would be unexpected from
an isotropic fireball in thermal equilibrium; however, it has previously been observed
that transverse expansion or flow 
results in increased ``temperature'' parameters for heavier states.  
Polleri {\it et al.} have shown~\cite{Polleri} that the heavier nuclear states give 
significantly more information
on the density profile and flow velocity than can be derived from lighter states ($\pi,K,p$).
Detailed comparisons between our data and such models should constrain our picture of the
flow evolution.

\begin{figure}[htb]
\begin{minipage}[b]{.46\linewidth}
\centering\epsfxsize=3in \epsfysize=3in \leavevmode \epsfbox[10 154 534 654]{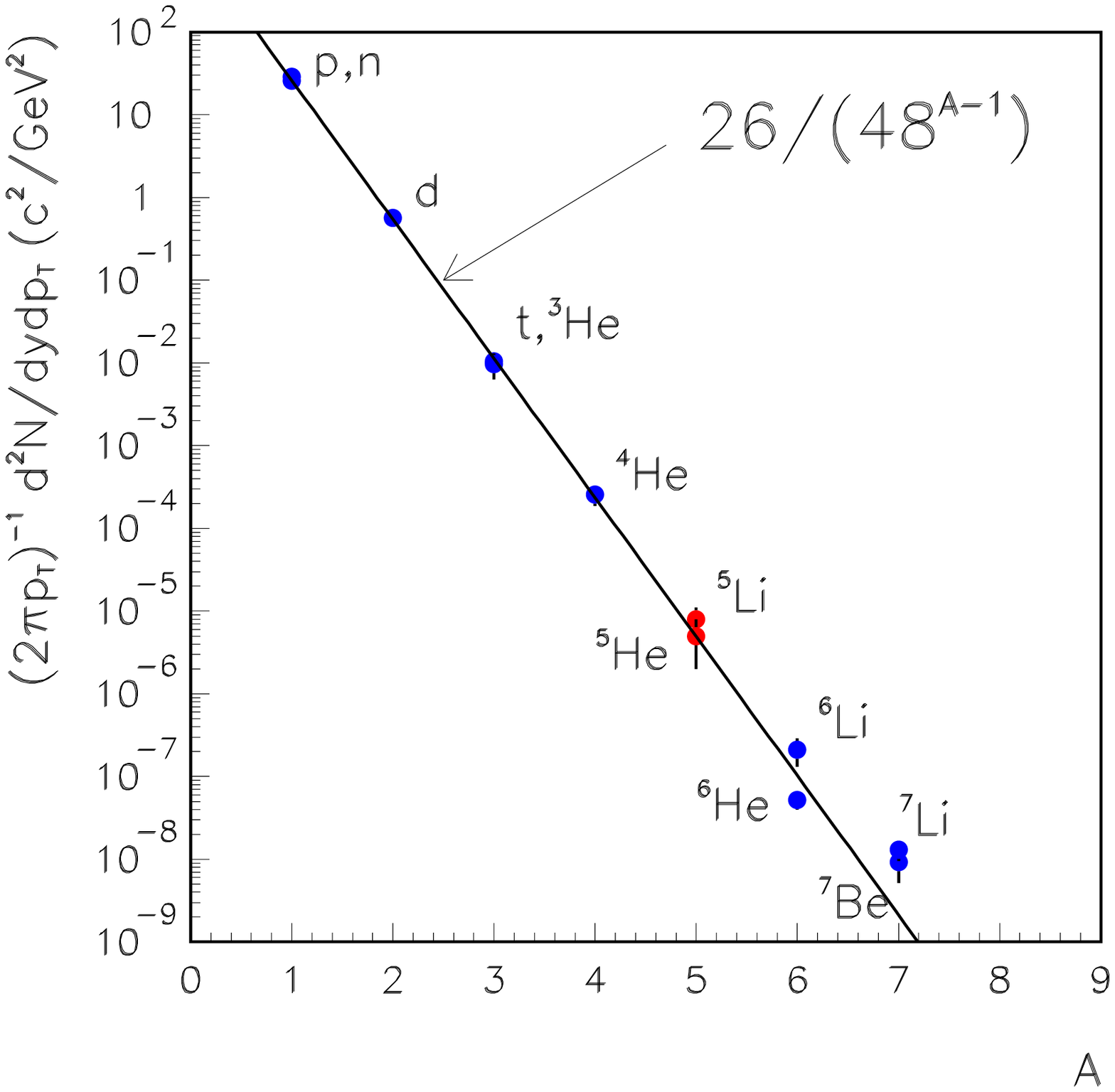}
\vspace{-0.45in}
\caption{Invariant yields per central collision at or near $y=1.9$, $p_{T}/A=200~MeV/c$ as a function of mass number $A$.}
\label{fig:nuclei_scaling}
\end{minipage}
\hspace{\fill}
\begin{minipage}[b]{.46\linewidth}
\centering\epsfxsize=3in \epsfysize=3in \leavevmode \epsfbox[10 154 534 654]{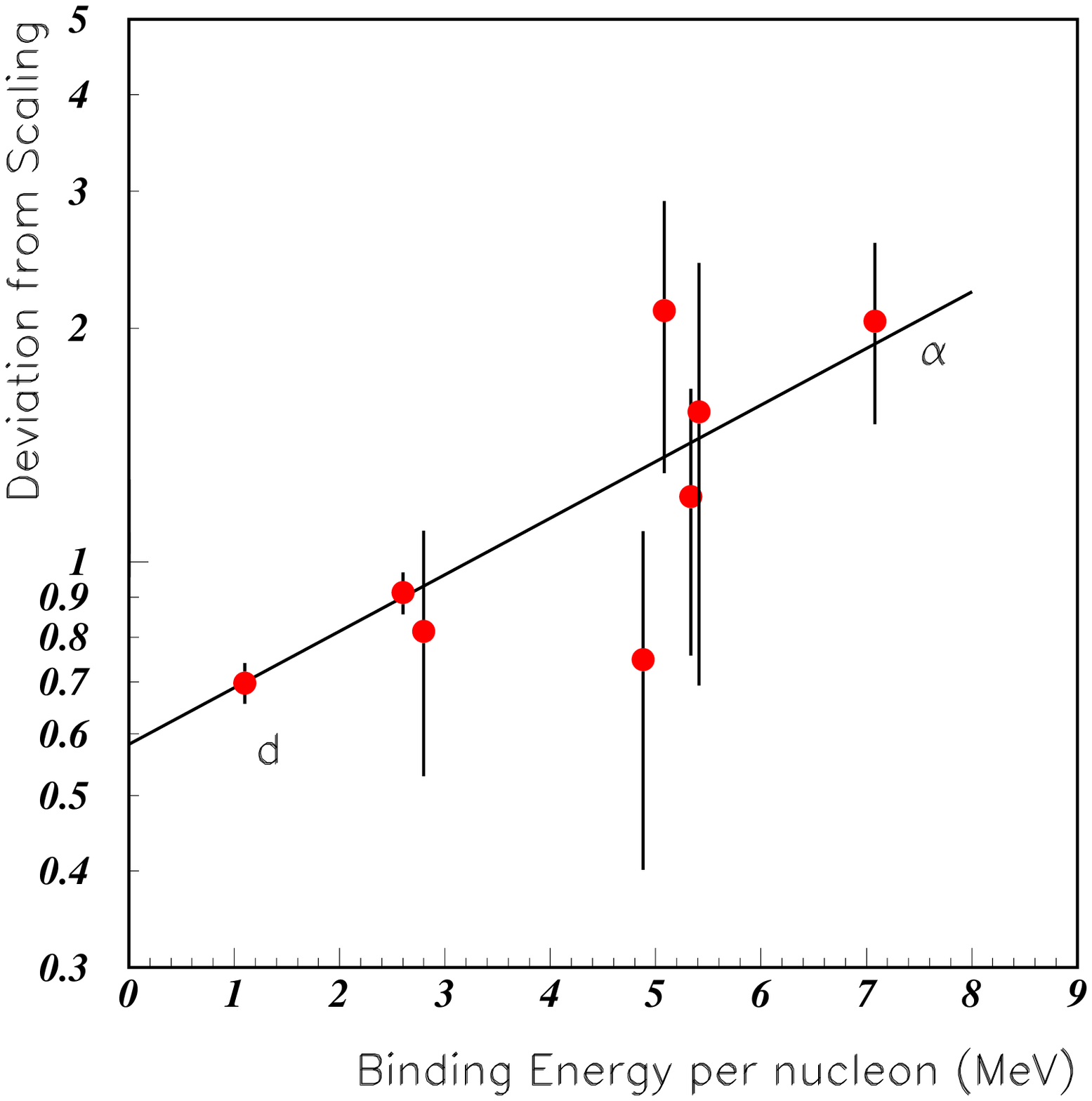}
\vspace{-0.45in}
\caption{Relative deviations from scaling relation after
correction for baryon number $A$, isospin $I_{z}$ and spin $J$ as a function of binding energy per nucleon (MeV).}
\label{fig:nuclei_binding}
\end{minipage}
\end{figure}

Using the high-mass trigger, we are also able to measure the invariant yields of light nuclei which
are produced down to the level of a few per billion central interactions.  We show in 
Figure~\ref{fig:nuclei_scaling} the measured yields of light 
nuclei up to baryon number A=7 as a function of
baryon number.  We observe a striking exponential behavior where the yields are characterized
by a penalty factor of approximately 48 per additional nucleon in the cluster.  It is 
interesting to note that in the early universe the lack of a stable A=5 state
has a large impact on the higher mass cluster yields.  However, since $^{5}Li$ and $^{5}He$ have
lifetimes longer than the entire time evolution of the heavy ion collision, their yields appear
to roughly scale with the other stable nuclei.  We are able to measure these states through a two
particle reconstruction of the $^{4}He$ and $p$ or $n$ and appropriate background subtraction.

Although the data are reasonably described by this simple scaling over ten order of magnitude
in yield, there are deviations.  There are some easily understood
deviations from scaling due to differences in isospin and spin factors.
In addition, as shown in Figure~\ref{fig:nuclei_binding}, it appears that there is a  
dependence on the binding energy per nucleon of the nuclear state.  
The weakly bound deuteron appears to have a suppressed yield 
relative to the most bound state that we measure, the alpha particle.  This dependence
on how tightly bound the object is may be an autocorrelation with the dependence
on the nuclear size.

\vspace{-0.05in}
\section{ANTINUCLEI} 

The production of antibaryons has long been recognized as a powerful tool for the
study of the baryon dense heavy ion collision environment.  At AGS energies
antiproton production is suppressed in initial collisions and
is further reduced due to their high annihilation cross section.  Antideuterons
are actually below threshold in nucleon-nucleon collisions at the 11.6 A GeV/c 
applicable for our experiment.  Thus, it is expected that antideuterons are formed from
the coalescence of antiprotons and antineutrons at the freeze out surface.

\begin{figure}[htb]
\begin{minipage}[b]{.46\linewidth}
\centering\epsfxsize=3in \epsfysize=3in \leavevmode \epsfbox[10 154 534 654]{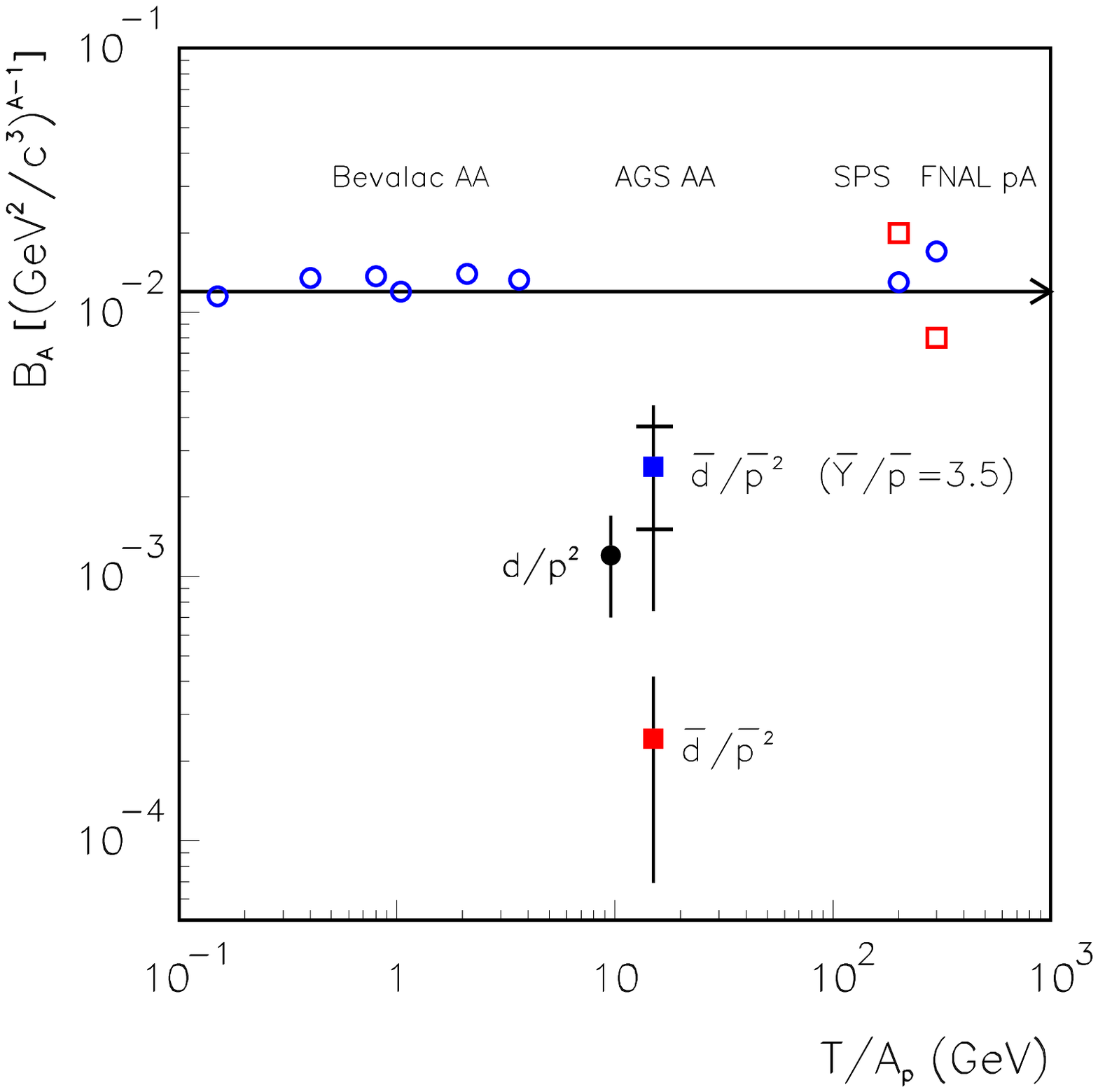}
\vspace{-0.45in}
\caption{Preliminary coalescence scale factor $\overline{B_{2}}$ for antideuterons using our 
measured antiproton yield 
and also using the antiprotons removing the contribution from strange antibaryons.  The 
corrected value agrees with our measured value for $B_{2}$ for deuterons. }
\label{fig:dbar}
\end{minipage}
\hspace{\fill}
\begin{minipage}[b]{.46\linewidth}
\centering\epsfxsize=3in \epsfysize=3in \leavevmode \epsfbox[10 146 556 689]{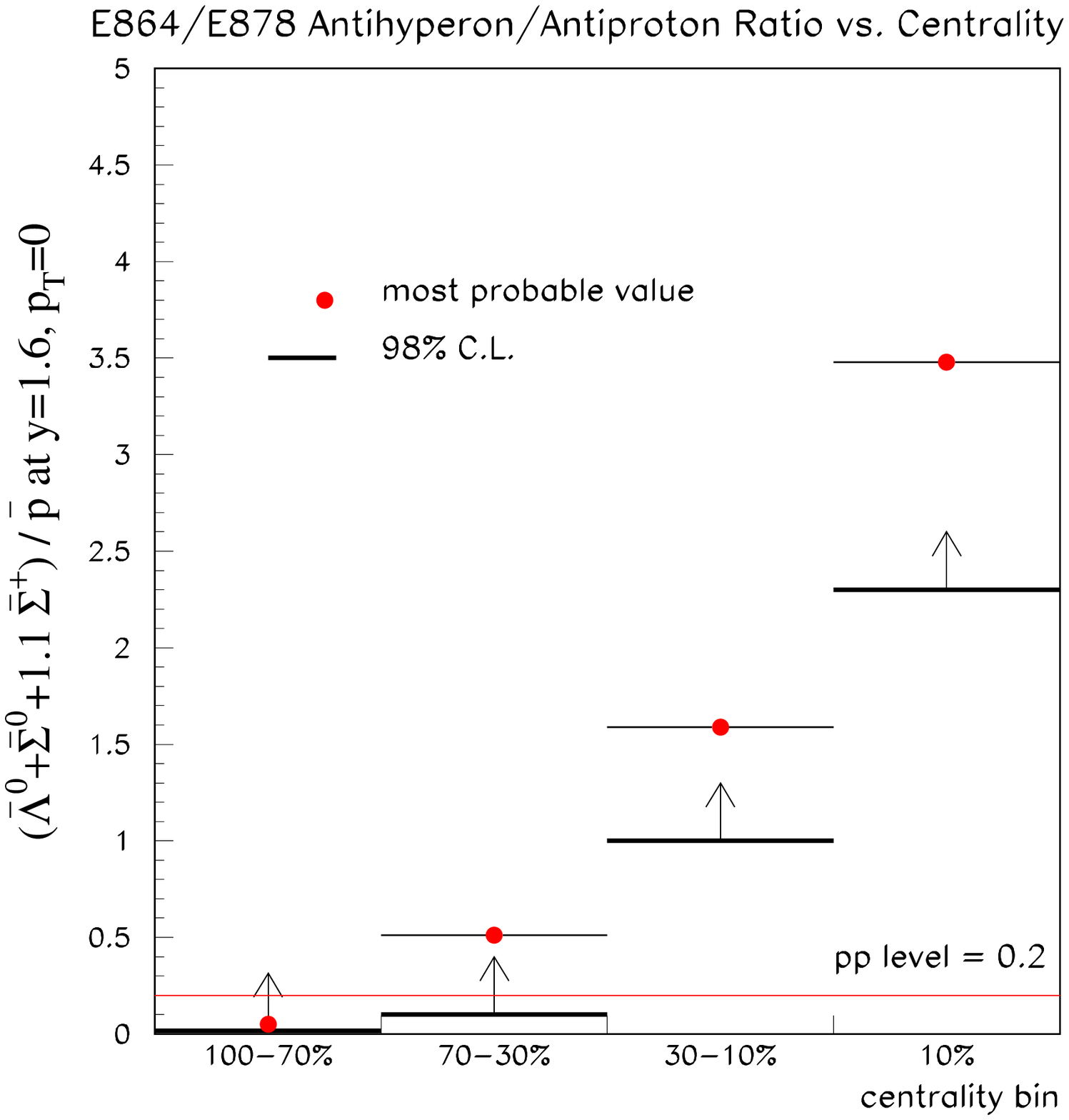}
\vspace{-0.45in}
\caption{The inferred ratio of strange antibaryons to antiprotons at midrapidity and
low transverse momentum as a function of collision centrality.  
These results are calculated from the antiproton measurements from E878 and E864.}
\label{fig:ybar}
\end{minipage}
\end{figure}

It has been theorized that due to the large absorption, antibaryons are emitted
only on the surface of the collision volume.  This unusual emission surface results
in a suppressed yield of antinuclei~\cite{stanislaw}.  We present here the first
measurement of antideuterons in $Au$ induced collisions at the AGS.  
We show in Figure~\ref{fig:dbar} the coalescence scale 
factor $\overline{B_{2}}$ (the ratio of antideuterons to antiprotons squared)
which appears to be almost an order of magnitude below the rate for deuterons $B_2$; 
thus leading one to conclude that surface emission may be a correct hypothesis.

However, we have previously published a surprising result on the yield of strange
antibaryons relative to antiprotons~\cite{ybar}.  E864 measures antiprotons which include
those that result from the decay of strange antibaryons ($\overline{\Lambda}$,
$\overline{\Sigma}$,$\overline{\Xi}$, etc.).  
When we compare our results to those of experiment E878
which measures only primordial antiprotons, we observe a statistically significant
difference.  This difference can be attributed to the decay products of strange
antibaryons.  Thus, we have calculated the ratio of strange antibaryons to
antiprotons near midrapidity and low transverse momentum (in the region of the
experimental acceptance) as a function of centrality, shown in Figure~\ref{fig:ybar}.
The implied ratio of $\overline{Y}/\overline{p}=3.5$ for the most central collisions is much above the
ratio from $p-p$ and $e^{+}-e^{-}$ collisions of $\overline{Y}/\overline{p}=0.2-0.4$.
This large ratio appears to be confirmed by preliminary E917 results shown at this
conference~\cite{hofman}.

If we consider that, in fact, most of the antiprotons included in the denominator of the
antideuteron scale factor are really strange antibaryons, then we have
underestimated the $\overline{B_2}$ value.  Since strange antibaryons decay long after the
coalescence process has taken place, they do not contribute to the yield
of antideuterons.  If we correct the $\overline{B_2}$ value for these contributions, then
it increases the scale factor by over an order of magnitude and into agreement
within statistical and systematic errors of the deuteron scale factor.
Both values are below the $B_2$ values from low energy $A-A$ collisions and high 
energy $p-p$ collisions, indicating a much larger freeze out volume for this system.

\vspace{-0.05in}
\section{HYPERNUCLEI}
\begin{figure}[htb]
\begin{minipage}[b]{.46\linewidth}
\centering\epsfxsize=3in \epsfysize=3in \leavevmode \epsfbox[10 154 534 654]{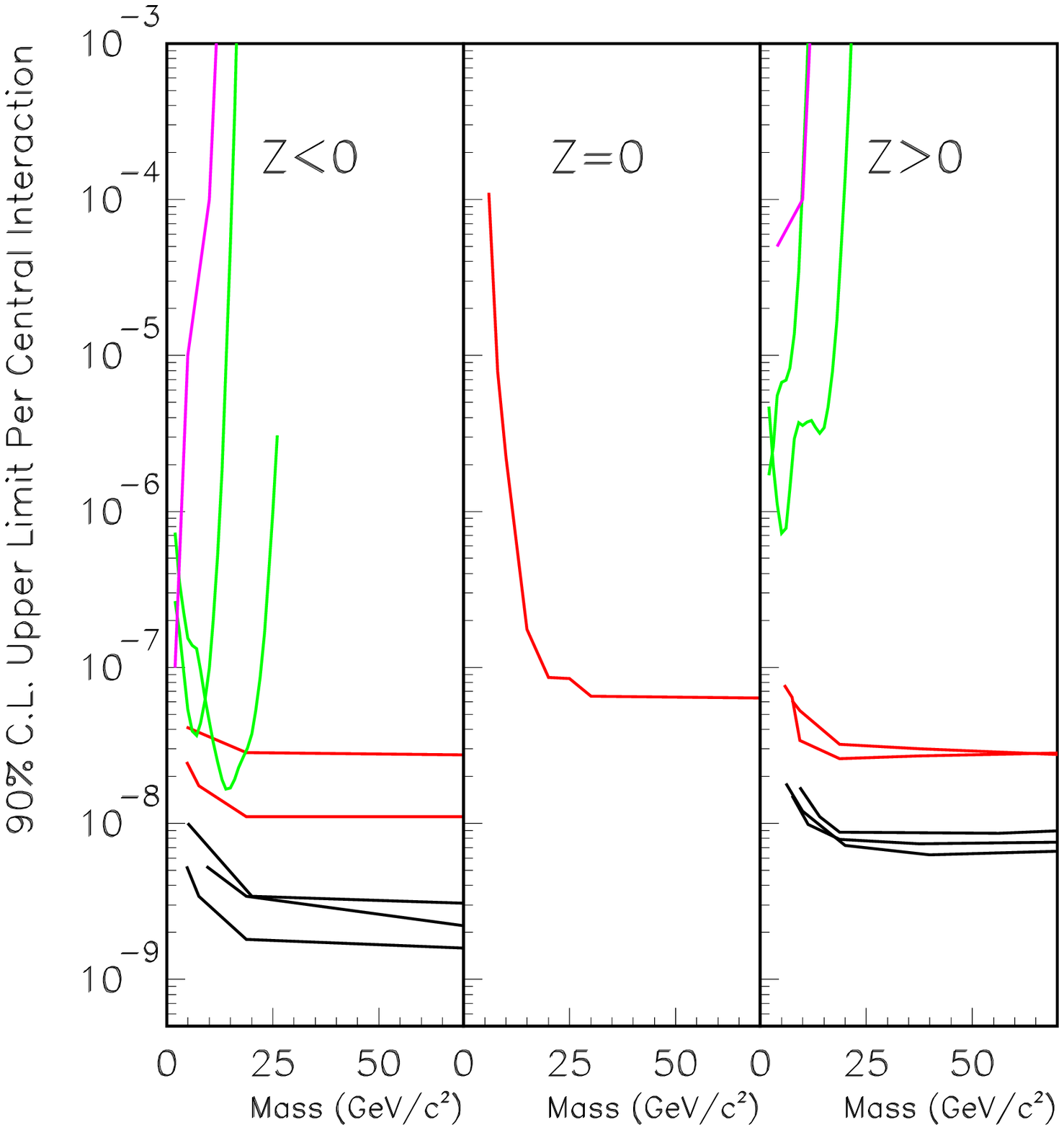}
\vspace{-0.45in}
\caption{90\% Confidence Level Upper Limits on the production of strange quark matter (negative,
positive and neutral charged states).  The upper curves at low mass are from E878 and E886.  The 
lower gray curves are the E864 published results and even lower black curves are the preliminary
E864 final results.}
\label{fig:sqm}
\end{minipage}
\hspace{\fill}
\begin{minipage}[b]{.46\linewidth}
\centering\epsfxsize=3in \epsfysize=3in \leavevmode \epsfbox[10 154 534 654]{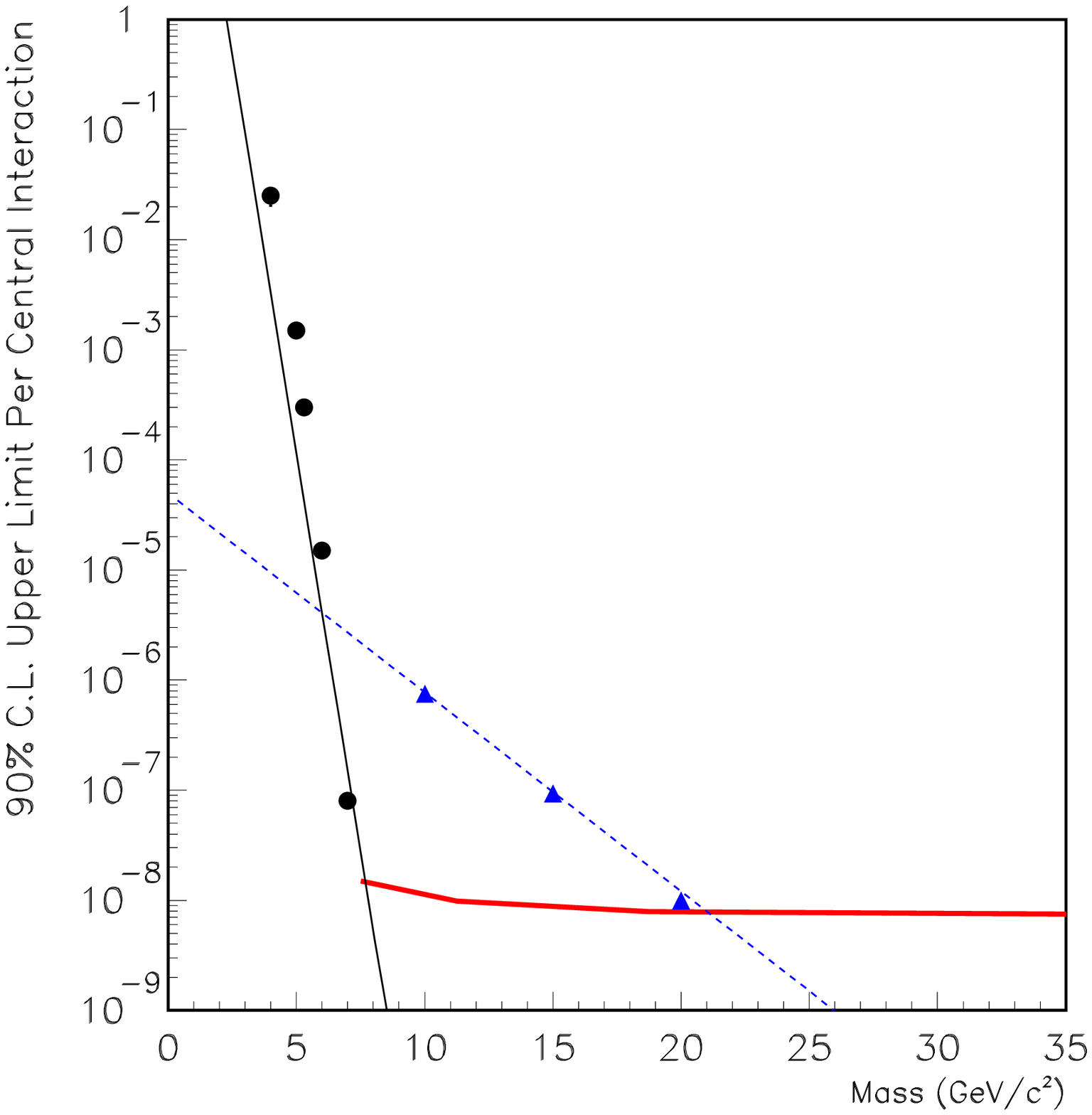}
\vspace{-0.45in}
\caption{90\% Confidence Level Upper Limits on the production of $Z=+2$ SQM.  Also shown are
predictions for their production in a cooling quark plasma (triangles) and via hypernuclei
acting as doorway states (circles).}
\label{fig:model}
\end{minipage}
\end{figure}

Unlike the early universe, the heavy ion system
freezes out on a time scale 10-20 fm/c, on the order of the strong interaction.  
Therefore, there are a substantial number of strange baryons which have
not decayed weakly that
contribute to a ``strange'' nucleosynthesis process.  
Here we have searched for the state $^{3}_{\Lambda}H$ via its
decay mode to $^{3}He$ + $\pi$.  We have a preliminary analysis of 2/3 of the 
relevant data sample and observe a signal at approximately the 2$\sigma$ level.
We measure a preliminary invariant yield of $^{3}_{\Lambda}H$ in the kinematic
range $1.6<y<2.8$ and $0<p_{T}<2~GeV$ of $2.6 \pm 1.4 \times 10^{-4}~c^{2}/GeV^{2}$.
Dover {\it et al.} have estimated that there
is an additional penalty factor for adding a strange baryon (instead of non-strange)
to a nucleus of 1/3 due to the smaller abundance of strange baryons~\cite{dover}.  Using 
the above measured invariant yield of $^{3}_{\Lambda}H$ combined with our measurements
of p, n, $^{3}He$, and the E891 measure of $\Lambda$~\cite{lambda}, 
we can check this model.
Our preliminary results show that in addition to the penalty for the smaller
abundance of $\Lambda$, there is an additional suppression of $0.162 \pm 0.088$.
This factor indicates that it is more difficult to make strange clusters than
originally predicted.   This suppression factor may imply that production of 
the $H$ di-baryon and other strange quark matter states
via coalescence will also be suppressed.

\section{STRANGE QUARK MATTER}
It has been theorized that strange quark matter (SQM), composed of nearly equal
numbers of up, down and strange quarks, might be meta-stable or even completely
stable in bulk.  Heavy ion collisions are
the best known environment for the production of small states of strange quark
matter.  The collisions produce a large number of strange quarks in a relatively small
configuration space.  In addition, the possible transition to a deconfined quark-gluon plasma 
is a natural environment to give rise to such new quark composites.  

We have not observed any evidence for meta stable (lifetimes of order 50 ns or
greater) charged or neutral $Z=\pm 0,1,2,3,>3$ SQM states.  We have
previously published 90\% confidence level upper limits based on data taken in 
1995~\cite{sqm_prl,sqm_npa,sqm_prc}.  Here we present our final limits on charged
states using our complete data sample.  Shown in Figure~\ref{fig:sqm} are the
upper limits from Experiments E878 and E886
in addition to our published results (grey lines).   Our final limits (black lines)
show that we rule out the production of SQM at the level of a $2-9 \times 10^{-9}$
per 10\% central collision. 

With the completion of the fixed target program at the BNL-AGS and the soon to be 
published final limits from Experiment E864, it is an appropriate time to ask what is the
significance in the absence of any observation.  Do these results imply
that SQM in the mass range $A=3-100$ is not meta
stable?  In order to answer this question, we must examine the various
production models for the creation of strangelets in these reactions.

Our results rule out a number of published predictions\cite{crawford} for SQM production in
a plasma scenario as shown in Figure~\ref{fig:model}; however,
these model calculations are not well constrained.
There are also models which predict the rate of SQM production assuming 
that multi-strange hypernuclei are formed via
coalescence at hadronic freeze out.  If a SQM state with
the same quantum numbers is more stable than this hypernucleus, the
hypernucleus may act as a ``doorway'' and transition to the quark matter
state.  
Using the number from~\cite{dover}, our experiment is sensitive to states of 
baryon number $A=6-7$ and strangeness $S=2-3$.  However, in light of our
preliminary measure of hypernuclei production, this would appear to overestimate
our sensitivity.

\vspace{-0.02in}
\section{SUMMARY}
E864 has completed the most sensitive search in relativistic 
heavy ion collisions at the AGS for exotic nuclear states and
novel quark composites (strange quark matter) to date.  We observe no SQM and thus
set the most stringent upper limits on its production.  We have 
extended the study of coalescence to a new regime including nuclei up to $A=7$,
unstable nuclei $^{5}He,^{5}Li$, hypernuclei $^{3}_{\Lambda}H$, and 
antideuterons.  The overall scaling of nuclear yields, the suppression of
hypernuclei, and the extreme enhancement of strange antibaryons present
a challenge to the theoretical community.
\vspace{-0.02in}
\section{ACKNOWLEDGMENTS}
We gratefully acknowledge the efforts of the AGS staff and support
from the DOE High Energy and Nuclear Physics Divisions and the NSF.


\end{document}